\documentclass[runningheads]{llncs}
\usepackage[T1]{fontenc}
\usepackage{graphicx}
\usepackage{orcidlink}
\usepackage{amsmath}
\usepackage{amssymb}
\begin{document}
\title{Bangla Music Genre Classification Using Bidirectional LSTMS}
\author{Muntakimur Rahaman\inst{1} \and
Md Mahmudul Hoque$^*$\inst{2}\orcidlink{0000-0002-2618-4157} \and
Md Mehedi Hassain\inst{3}\orcidlink{0009-0000-0818-245X}}
\authorrunning{M. Rahaman et al.}
\institute{Dept. of Computer Science \& Engineering, Bangladesh Army International University of Science and Technology, Cumilla Cantonment, Cumilla - 3501, Bangladesh \\
\email{muntakim.cse@gmail.com}
\and
Dept. of Computer Science \& Engineering, CCN University of Science and Technology, CCN Road, Cumilla - 3506, Bangladesh. \\ 
\email{cse.mahmud.evan@gmail.com$^*$}\\
\and
Dept. of Electrical \& Electronic Engineering, International Islamic University Chittagong,Chattogram - 4318, Bangladesh. \\
\email{mdmehedihasan225588@gmail.com}\\
}
\maketitle              % typeset the header of the contribution
\begin{abstract}
Bangla music is enrich in its own music cultures. Now a days music genre classification is very significant because of the exponential increase in available music, both in digital and physical formats. It is necessary to index them accordingly to facilitate improved retrieval. Automatically classifying Bangla music by genre is essential for efficiently locating specific pieces within a vast and diverse music library. Prevailing methods for genre classification predominantly employ conventional machine learning or deep learning approaches. This work introduces a novel music dataset comprising ten distinct genres of Bangla music. For the task of audio classification, we utilize a recurrent neural network (RNN) architecture. Specifically, a Long Short-Term Memory (LSTM) network is implemented to train the model and perform the classification. Feature extraction represents a foundational stage in audio data processing. This study utilizes Mel-Frequency Cepstral Coefficients (MFCCs) to transform raw audio waveforms into a compact and representative set of features. The proposed framework facilitates music genre classification by leveraging these extracted features. Experimental results demonstrate a classification accuracy of 78\%, indicating the system's strong potential to enhance and streamline the organization of Bangla music genres.
\keywords{Bangla music genre classification \and Bidirectional LSTM \and Digital music indexing \and MFCC \and Music feature extraction.}
\end{abstract}
\section{Introduction}
Music holds a vital place in daily life, shaping emotions, culture, and decision-making. As one of humanity's most universal and influential art forms, it serves as a powerful medium for emotional expression, cultural preservation, and social connection, capable of transforming moods and bringing solace during difficult times. The music industry, valued at billions of dollars, plays a significant role in advertising and entertainment, where its psychological impact drives consumer behavior and enhances storytelling. However, the very attributes that make music so powerful—its rhythmic complexity, melodic diversity, and lyrical depth—also pose significant challenges for automated categorization. Accurate music classification is crucial for creating personalized recommendation systems but remains challenging due to music's diverse attributes including rhythm, melody, harmony, and lyrics. While genre classification research for English music is well-developed, Bangla music, known for its rich mix of classical and modern styles, has seen limited exploration in this area. Bengali is the seventh most common language globally, with a speaker base of roughly 245 million located chiefly in Bangladesh and Eastern India\cite{islam2009research,hossain2005review}. for its perceived lyrical and melodic quality and phonetic richness, Bengali is often considered the sweetest language in the world. The phonological properties of Bengali, frequently characterized by euphonic vowel clusters and rhythmic syllabic patterns, provide an exceptional foundation for Bangla music's characteristically seamless text-setting and melodic fluidity. Several approaches have been conducted to address music classification challenges. Researchers have frequently utilized established machine learning algorithms, notably SVM and KNN, to classify musical genres by analyzing acoustic features derived from MFCCs\cite{ahmed2022machine,hasan2021bangla,rahman2012bangla,rokib2023machine,khan2024analyzing,humayra2024music} . However, these approaches often fail to capture the complex relationships within musical data. Modern deep learning techniques, including Deep Neural Networks (DNN), Convolutional Neural Networks (CNN), and Recurrent Neural Networks (RNN), offer superior accuracy by effectively modeling intricate patterns and temporal dependencies.

Our study aims to develop a robust and computationally efficient framework for Bangla music genres by investigating three key aspects: the unique acoustic characteristics of Bangla music, the effectiveness of various neural network approaches for genre classification, and methods to overcome data scarcity challenges specific to Bangla music. We particularly focus on how the linguistic properties of Bengali influence musical features and classification performance.

This research seeks to bridge the gap in Bangla music classification by leveraging advanced neural network models to categorize Bangla music genres. Enhancing the accuracy of genre classification can significantly improve recommendation systems and benefit the Bangla music industry. However, the study faces challenges such as limited labeled datasets, complex feature extraction processes, and resource constraints. Addressing these issues could greatly enhance the representation and accessibility of Bangla music in the digital space.

\section{Literature Review and Working Process}

This research focused on two primary tasks: feature engineering and the classification of Bangla songs. The dataset consisted of 10 genres: Bangla hip-hop, Bangla metal, Bangla rock, deshattobodhok, Palligiti, lalon giti, Nazrul Sangeet, Rabindra Sangeet, folk, and hamdanaat. The songs, originally in mp3 format, were converted to wav format. Features were extracted using techniques such as ZCR, spectral centroid, and MFCC. MFCCs, in particular, played a key role in Music Information Retrieval (MIR) and genre classification. To execute the classification task, a Bidirectional Long Short-Term Memory model was utilized, with its output fed into a subsequent dense layer for generating the final class labels.

\subsection{Related Works} 

Genre classification has a long and rich history, evolving from early methods of analyzing vocal and instrumental music in ancient times to the advanced digital techniques developed in the 20th century. The emergence of digital music platforms, such as iTunes, created a demand for enhanced user experiences, particularly through the development of recommendation systems. This need has driven extensive research into automated genre classification, leading to significant advancements in the field.
Kris West et al. \cite{inproceedings} made notable contributions by emphasizing the importance of onset detection for feature extraction in their work on the automated genre recognition in audio signals. A novel framework was introduced, leveraging unsupervised decision trees in conjunction with either linear discriminant analysis or a two-part Gaussian classifier.
Their results demonstrated that this classifier's topology provided a significant advantage over traditional Gaussian-based schemes by enabling the modeling of more complex data distributions. This work laid the foundation for many subsequent studies in the field. S. Patil et al. \cite{shete2014zero} focused on separating voiced and non-voiced speech segments using Zero Crossing Rate (ZCR) and energy-based features. By dividing speech samples into segments and analyzing ZCR and energy, they found that voiced segments typically exhibit a reduced ZCR and prominent energy, while unvoiced segments show a high frequency of zero crossings and diminished energy. This approach provided a reliable method for distinguishing between these two types of speech components, contributing to the broader field of audio analysis. Muhammad Ami Ali et al. \cite{Ali2017} explored the use of machine learning algorithms, specifically k-Nearest Neighbor (k-NN) and Support Vector Machine (SVM), in conjunction with Mel-Frequency Cepstral Coefficients (MFCC) for genre classification. They compared classifier performance with and without Principal Component Analysis (PCA) for dimensionality reduction. Their results indicated that both k-NN and SVM performed more accurately without dimensionality reduction, with SVM achieving an overall accuracy of 77\%. This study highlighted SVM's effectiveness as a classifier for music genre classification. Ali et al. \cite{aliYildiz} further advanced the field by applying signal processing methods and machine learning algorithms to the GTZAN dataset. Their work demonstrated the potential of combining traditional signal processing methods with modern machine learning approaches for accurate genre classification. Ajay et al. \cite{ajaysriram} contributed to this area by employing Logistic Regression, SVM, and k-NN to predict music genres based on features such as MFCC and tempo. Their research underscored the effectiveness of combining traditional machine learning models with audio features for genre classification tasks. Hareesh et al. \cite{hareesh} utilized Convolutional Neural Networks (CNNs) with spectrograms for genre classification. They trained four traditional machine learning classifiers and compared their performance, identifying the most impactful features for multi-class classification. Their evaluation on the Audio Set corpus yielded an area under the curve (AUC) of 0.894 for a combined model integrating their two proposed methods, demonstrating the strong capability of deep learning in this area. An analysis \cite{hoque2024comparison} of The Daily Star and Prothom Alo found Spectral Clustering most effective for news data segmentation, achieving Silhouette Scores of 89\% and 83\%.
Thomas et al. \cite{thomas} employed parallel CNNs for mood and genre identification using Mel-Spectrograms. Their methodology featured a sequential CNN and a dual-stream parallel CNN. The parallel model's separate channels independently learned time-varying and spectral features before their outputs were combined. This approach demonstrated the effectiveness of parallel architectures in capturing diverse aspects of musical data. Chun Pui Tang et al. \cite{10.1117/12.2501763} presented a hierarchical Long Short-Term Memory (LSTM) model for genre classification. They classified music into "strong" genres (e.g., hip-hop, metal, pop, rock, reggae) and "mild" genres (e.g., jazz, disco, country, classical, blues) based on musical intensity. Their hierarchical divide-and-conquer approach achieved an average classification accuracy of 50.00\% for 10-genre classification, outperforming single CNN-based methods. This work highlighted the benefits of hierarchical models in improving classification accuracy. 
A recent study explored \cite{hoque2023analyzing} forecasting monthly prices of rice, lentils, and wheat flour across divisions using various time series models. RNN and a Naïve Ensemble approach yielded the most accurate results based on MAE.
Bi Li et al. \cite{Li2016/11} discussed feature extraction techniques for music recommendation systems, focusing on temporal intensity, spectral magnitude, Mel-Frequency Cepstral Coefficients (MFCCs), and the spectral contour. These attributes were consolidated into a combined feature set, and the fractal complexity was quantified via the Hilbert transformation. Their methodology highlighted the structural recursion across global and sectional segments of audio, establishing a robust technique for music information retrieval. The results demonstrated the potential of fractal dimension-based feature extraction in enhancing music recommendation systems.
The existing body of research demonstrates a wide array of creative methodologies crafted for the purpose of music genre categorization. The domain has undergone substantial progression, moving from conventional signal analysis and statistical learning methods to sophisticated neural network frameworks. These developments have not only enhanced classification precision but have also established a foundation for subsequent innovations in music retrieval and recommendation technologies. Together, these studies underscore the critical roles played by feature representation, model architecture, and structured learning in tackling the intricacies of this multifaceted problem.

\section{Proposed Algorithm} \label{sec3}
\begin{figure}
    \centering
    \includegraphics[width=0.5\linewidth]{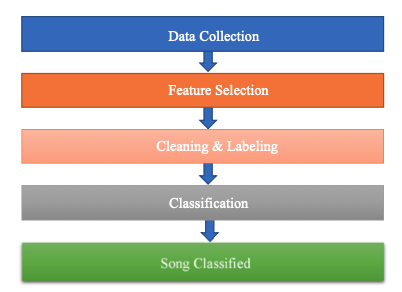}
    \caption{Music Genre Classification Model}
    \label{fig:methodology}
\end{figure}

An algorithm comprises a sequence of instructions designed to accomplish a specific task. For the categorization of Bangla music genres, we proposed an efficient deep learning algorithm. This involves feeding relevant data to the algorithm, allowing it to identify patterns and learn unique features through iterations with a specific learning rate. Over time, the algorithm self-improves in classification. Our methodology utilizes a widely recognized recurrent neural network architecture known as bidirectional long short-term memory (Bi-LSTM). This configuration allows the model to process sequential data in both temporal directions, significantly improving the precision of genre classification for Bangla music, as illustrated in Fig \ref{fig:methodology}.

\subsection{LSTM(LONG SHORT TERM MEMORY)}\label{subsec3}
LSTM, introduced by Hochreiter and Schmidhuber in the 1990s, with a forget gate added by Gers et al. in 1999 \cite{forgetgate}, addresses RNNs' short-term memory issues by modeling long-term dependencies. It gained prominence in 2009 after winning the ICDAR handwriting competition \cite{handwriting} and powers applications like translation, speech recognition, and time-series analysis in companies like Google and Facebook.   
With input, forget, and output gates, LSTMs manage sequential data effectively. Bidirectional LSTMs, introduced by Schuster and Paliwal \cite{650093}, enhance context by processing data in both directions, excelling in speech, text, and forecasting tasks. Reasons to use Bidirectional Long Short Term Memories:
\begin{itemize}
    \item LSTMs are better Recurrent Neural Network than GRU.
    \item Bidirectional LSTMs are used for entity extraction.
    \item It provides better result because of its memory property.
    \item Bidirectional LSTMs combined with dense network which helps classify Bangla music genres with better accuracy.
\end{itemize}

\subsection{Feature Engineering}
The music collection encompassed ten unique genres of Bangla music, such as ‘Bangla hip-hop’, ‘Bangla Metal’, ‘Bangla Rock’, ‘Deshattobodhok’, ‘Palligiti’, ‘Lalon Giti’, ‘Nazrul Sangeet’, ‘Rabindra Sangeet’, ‘Folk’, and ‘Hamdnaat’. The original audio files, stored in mp3 format, were first converted into wav format. As a standard audio file type, the wav format stores sound as waveform data, rendering it appropriate for computational analysis. To make the audio compatible with the model, relevant features were derived from the signals using established extraction methods. A range of acoustic descriptors were considered, including Zero Crossing Rate (ZCR), Spectral Centroid, Spectral Roll-off, Spectral Bandwidth, Chroma Features, Root-Mean-Square (RMS) Energy, Delta Coefficients, and Mel-Frequency Cepstral Coefficients (MFCCs). MFCCs were selected for this study owing to their proven capacity to effectively model the short-term power spectrum of audio. The technique applies a cosine transform to the logarithm of the spectral energy distributed on a perceptually motivated mel frequency scale, which has established MFCCs as a robust and widely adopted tool in Music Information Retrieval (MIR). The MIR application, particularly in genre classification, benefits significantly from MFCC. The MFCC coefficients were extracted from each song, and the process is illustrated in Fig \ref{fig:mfcc}. This approach ensured a robust and efficient feature extraction process, aligning with the requirements of the genre classification task.

\subsection{Model Architecture}
\begin{figure*}[htbp]
    \centering
    \begin{minipage}[b]{0.45\linewidth}
        \centering
        \includegraphics[width=\linewidth]{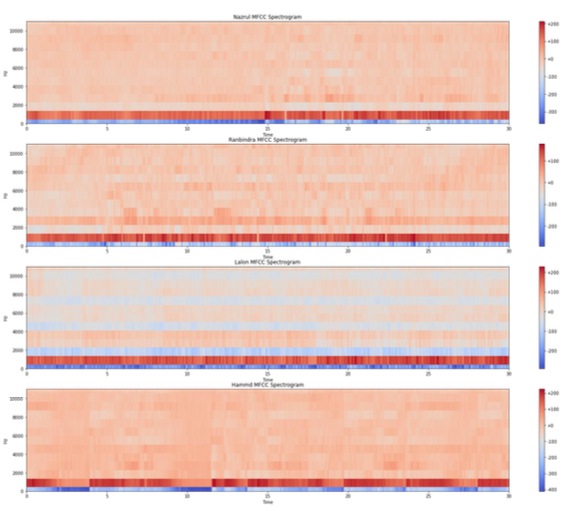}
        \caption{MFCC of 4 random songs from 4 genres.}
        \label{fig:mfcc}
    \end{minipage}
    \hfill
    \begin{minipage}[b]{0.45\linewidth}
        \centering
        \includegraphics[width=\linewidth]{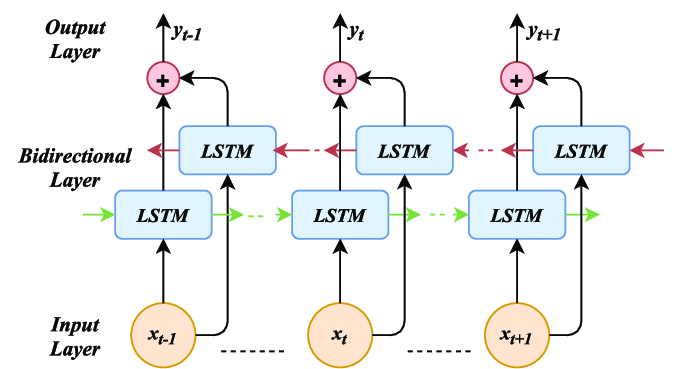}
        \caption{Model Architecture of Proposed Model.}
        \label{fig:architecture}
    \end{minipage}
\end{figure*}

The model is designed with a linear stack of layers, integrating several Bidirectional Long Short-Term Memory (Bi-LSTM) units along with batch normalization and fully-connected layers. The Bi-LSTM components analyze sequence data concurrently in chronological and reverse-chronological order, enabling the capture of intricate time-dependent structures that are fundamental to audio signal interpretation. The system processes sequential input data {x\_t-1, x\_t, x\_t+1} representing acoustic features (e.g., MFCCs) through parallel forward and backward LSTM layers, enabling comprehensive context capture from both past and future states. This bidirectional processing is particularly effective for music classification as it can identify characteristic patterns in rhythm (e.g., taal cycles) and melody that may span multiple time steps. The output layer generates predictions {y\_t-1, y\_t, y\_t+1} for frame-level analysis or combines final hidden states for sequence-level classification. This approach addresses key challenges in Bangla music analysis by simultaneously modeling the progressive evolution of musical elements and the inherent bidirectional relationships between lyrical and instrumental components, while overcoming the vanishing gradient problem through LSTM's gating mechanisms. The architecture of the model is illustrated in the Fig \ref{fig:architecture}, which provides a visual representation of the structure employed in this research. The model architecture is formally characterized by the following mathematical representations:

\begin{itemize}
    \item \textbf{Input Representation}
    \end{itemize}
    
Let $\mathbf{X} = (\mathbf{x}_1, \mathbf{x}_2, \ldots, \mathbf{x}_T)$ denote the input sequence, where each $\mathbf{x}_t \in \mathbb{R}^d$ represents a $d$-dimensional feature vector at time step $t$. For Bangla music, we compute:
\begin{equation}
\mathbf{x}_t = \left[\text{MFCC}(t), \Delta\text{MFCC}(t), \text{Chroma}(t)\right]^\top,
\end{equation}
where $\text{MFCC}(t)$ captures timbral features, $\Delta\text{MFCC}(t)$ their temporal derivatives, and $\text{Chroma}(t)$ harmonic content aligned with Bengali musical scales.

\begin{itemize}
    \item \textbf{Bidirectional LSTM Layer}
    \end{itemize}
The model processes $\mathbf{X}$ bidirectionally via:
\begin{align}
\overrightarrow{\mathbf{h}}_t &= \text{LSTM}_{\text{forward}}(\mathbf{x}_t, \overrightarrow{\mathbf{h}}_{t-1}), \\
\overleftarrow{\mathbf{h}}_t &= \text{LSTM}_{\text{backward}}(\mathbf{x}_t, \overleftarrow{\mathbf{h}}_{t+1}),
\end{align}
where $\overrightarrow{\mathbf{h}}_t, \overleftarrow{\mathbf{h}}_t \in \mathbb{R}^h$ are hidden states. The combined representation:
\begin{equation}
\mathbf{h}_t = \left[\overrightarrow{\mathbf{h}}_t; \overleftarrow{\mathbf{h}}_t\right] \in \mathbb{R}^{2h}
\end{equation}
preserves contextual relationships. The LSTM gates are computed as:
\begin{align}
\mathbf{f}_t &= \sigma(\mathbf{W}_f[\mathbf{h}_{t-1}; \mathbf{x}_t] + \mathbf{b}_f) \quad \text{(forget gate)} \\
\mathbf{i}_t &= \sigma(\mathbf{W}_i[\mathbf{h}_{t-1}; \mathbf{x}_t] + \mathbf{b}_i) \quad \text{(input gate)} \\
\mathbf{o}_t &= \sigma(\mathbf{W}_o[\mathbf{h}_{t-1}; \mathbf{x}_t] + \mathbf{b}_o) \quad \text{(output gate)}
\end{align}

\begin{itemize}
    \item \textbf{Output Layer}
    \end{itemize}

For \textbf{frame-level} genre probabilities:
\begin{equation}
\mathbf{y}_t = \text{softmax}(\mathbf{W}_y \mathbf{h}_t + \mathbf{b}_y), \quad \mathbf{y}_t \in \mathbb{R}^{|G|},
\end{equation}
where $|G|$ is the number of genres. For \textbf{sequence-level} classification:
\begin{equation}
\mathbf{y} = \text{softmax}\left(\mathbf{W}_y \left[\overrightarrow{\mathbf{h}}_T; \overleftarrow{\mathbf{h}}_1\right] + \mathbf{b}_y\right).
\end{equation}
The model minimizes cross-entropy loss:
\begin{equation}
\mathcal{L}(\theta) = -\sum_{t=1}^T \sum_{i=1}^{|G|} y_t^i \log(\hat{y}_t^i),
\end{equation}
where $\theta = \{\mathbf{W}_h, \mathbf{U}_h, \mathbf{b}_h, \mathbf{W}_y, \mathbf{b}_y\}$ are trainable parameters.

\begin{itemize}
    \item \textbf{Musical Context Integration}
    \end{itemize}

To encode Bengali linguistic features:
\begin{equation}
\tilde{\mathbf{x}}_t = \left[\mathbf{x}_t; \mathbf{p}_t\right],
\end{equation}
where $\mathbf{p}_t \in \mathbb{R}^k$ is derived from lyrics using a pretrained Bengali phoneme model.

\section{Experimental Results}

\begin{table}[h]
\centering
\begin{tabular}{|c|c|}
\hline
\textbf{Model} & \textbf{Accuracy} \\ \hline
Logistic Regression & 64\% \\ \hline
SVM & 21\% \\ \hline
K-NN(n=10) & 45\% \\ \hline
ANN & 72\% \\ \hline
CNN & 65\% \\ \hline
LSTM & 74\% \\ \hline
Our Proposed Bi-LSTM & 78\% \\ \hline
\end{tabular}
\caption{Comparative study of different approaches with respective accuracy}
\label{tab1}
\end{table}

Numerous prior studies have investigated musical genre categorization by applying both conventional machine learning and modern deep learning methods across various languages. Frequently employed techniques in these studies comprise logistic regression, support vector machines (SVM), k-nearest neighbors (K-NN), artificial neural networks (ANN), convolutional neural networks (CNN), and long short-term memory (LSTM) networks.

In our experimental framework, we assessed the performance of Logistic Regression, Linear Regression, SVM, K-NN, and ANN models using our collected dataset. Given computational limitations—specifically those related to resource-demanding spectrogram creation and neural network training procedures—we opted to omit CNN and LSTM models from our primary evaluation. To maintain computational efficiency while preserving relevant audio information, we utilized Mel-Frequency Cepstral Coefficients (MFCCs) as our primary feature representation method.

\begin{figure}[h!]
    \centering
    \begin{minipage}{0.45\textwidth}
        \centering
        \includegraphics[width=\linewidth]{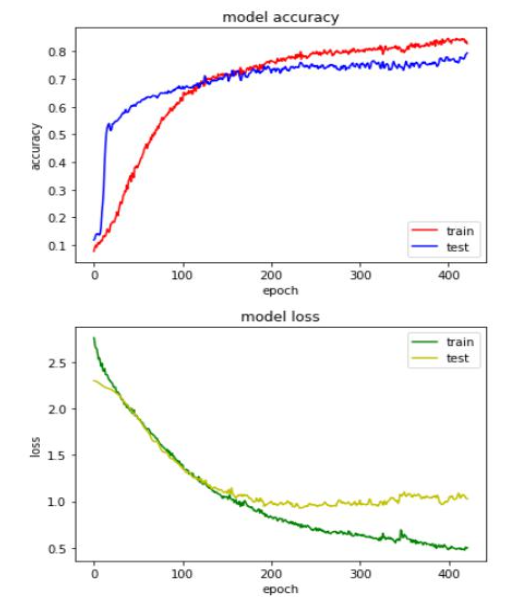}
        \caption{Accuracy and Loss graph per epoch.}
        \label{fig:accuracy}
    \end{minipage}\hfill
    \begin{minipage}{0.5\textwidth}
        \centering
        \includegraphics[width=\linewidth]{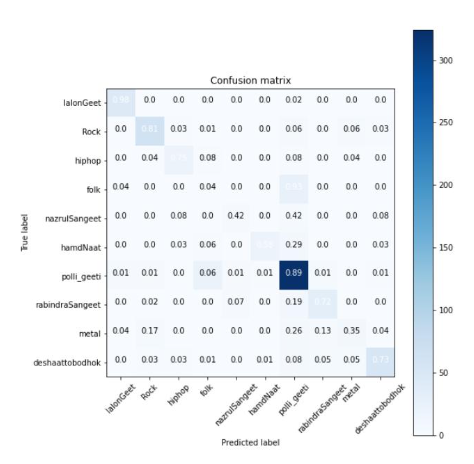}
        \caption{Confusion matrix of our model.}
        \label{fig:confusion}
    \end{minipage}
\end{figure}

As evidenced by prior research, neural networks typically outperform traditional machine learning models as dataset sizes increase. Our comparative analysis Table \ref{tab1} demonstrates this trend, with the proposed Bidirectional LSTM achieving 78\% accuracy, significantly surpassing conventional methods (e.g., SVM at 21\%, K-NN at 45\%). This performance advantage is further illustrated in Figure \ref{fig:accuracy}. These experimental outcomes justified the selection of our architecture for subsequent refinement and optimization.

\subsection{Model Evaluation with Metrics}

\begin{table}[]
    \centering
    \begin{tabular}{|c|c|c|c|}
    \hline
         & Precision & Recall & F1-Score \\
         \hline
        LalonGeet & 0.85 & 0.98 & 0.91 \\
        \hline
        Rock & 0.85 & 0.81 & 0.83 \\
        \hline
        hiphop & 0.72 & 0.75 & 0.73 \\
        \hline
        folk & 0.04 & 0.04 & 0.04 \\
        \hline
        nazrulSangeet & 0.45 & 0.42 & 0.43 \\
        \hline
        hamdNaat & 0.86 & 0.58 & 0.69 \\
        \hline
        polli\_geeti & 0.83 & 0.89 & 0.86 \\
        \hline
        rabindraSangeet & 0.72 & 0.72 & 0.72 \\
        \hline
        metal & 0.44 & 0.35 & 0.39 \\
        \hline
        deshaattobodhok & 0.88 & 0.73 & 0.80 \\
        \hline
    \end{tabular}
    \caption{Classification Report of our model.}
    \label{23}
\end{table}

After completing the training and validation phases, we evaluated the model using a separate testing dataset consisting of 718 samples. The model achieved a testing accuracy of 78.69\%, demonstrating its ability to generalize to unseen data. The training and testing accuracy curves along with loss metrics are visualized in Figure \ref{fig:accuracy}, demonstrating stable convergence during the learning process. Further insights into the model's classification behavior are revealed through the confusion matrix in Figure \ref{fig:confusion}, which highlights both the model's strengths and specific areas of misclassification across different musical genres.

A comprehensive classification report presented in Table \ref{23} details the model's performance metrics for each genre category. The results show particularly strong performance for Lalon Geet with an F1-score of 0.91 and Polli Geeti with an F1-score of 0.86, while indicating challenges with certain genres like Folk at 0.04 and Metal at 0.39. These evaluation indicators together deliver a comprehensive quantitative appraisal of the model’s forecasting proficiency, with precision indicating classification trustworthiness, recall assessing detection thoroughness, and the F1-score offering a harmonized measure of both.

\section{Conclusion}

This study investigates the automatic categorization of Bangla music genres through the application of Bidirectional Long Short-Term Memory (BiLSTM) neural networks, attaining a classification accuracy of 78\%. Utilizing Mel-Frequency Cepstral Coefficients (MFCCs) for acoustic feature representation, the work highlights the capability of deep learning approaches to streamline genre identification, thereby facilitating improved management and retrieval of musical archives. The outcomes hold tangible value for digital music services, broadcasting platforms, and streaming providers by supporting the development of refined recommendation mechanisms and enriched listener engagement. Furthermore, this research aids in the digital preservation and dissemination of Bangla musical heritage, supporting efforts in cultural conservation. Subsequent improvements in dataset scale, feature engineering, and algorithmic refinement are expected to increase the system’s performance and adaptability, fostering continued progress within the music technology domain.

\bibliographystyle{splncs04}
\bibliography{mybibliography}
%Author, F.: Article title. Journal \textbf{2}(5), 99--110 (2016)

%Author, F., Author, S.: Title of a proceedings paper. In: Editor,
%F., Editor, S. (eds.) CONFERENCE 2016, LNCS, vol. 9999, pp. 1--13.
%Springer, Heidelberg (2016). \doi{10.10007/1234567890}

%Author, F., Author, S., Author, T.: Book title. 2nd edn. Publisher,

%Author, A.-B.: Contribution title. In: 9th International Proceedings
%on Proceedings, pp. 1--2. Publisher, Location (2010)

%LNCS Homepage, \url{http://www.springer.com/lncs}, last accessed 2023/10/25
%\end{thebibliography}
\end{document}